\documentclass[12pt,oneside,reqno]{amsart}

\usepackage{geometry}
\usepackage{amsmath}
\counterwithout{footnote}{section}
\DeclareMathOperator{\Tr}{Tr}
\usepackage{setspace}
\usepackage{graphicx}

\usepackage{cleveref}
\usepackage{multirow}
\usepackage[normalem]{ulem}
\usepackage{amsmath,amssymb,amsfonts}
\usepackage{mathtools,leftindex,tensor,mhchem}
\usepackage{amsthm}

\usepackage{enumitem}
\usepackage{pifont}
\usepackage{mathrsfs}
\usepackage{bm}
\usepackage[title]{appendix}
\usepackage{xcolor}
\usepackage{textcomp}
\usepackage{accents}
\usepackage{manyfoot}
\usepackage{enumerate}
\usepackage{algorithm}
\usepackage{algorithmicx}
\usepackage{algpseudocode}
\usepackage{listings}

\newtheorem{definition}{Definition}

\usepackage{comment}

\def\eq#1{Eq.~$(\ref{#1})$}

\def\S{{\mathscr{S}}}

\def\({\textup{(}}
\def\){\textup{)}}

\begin{document}

\title[Minkowski Axioms]{Axioms for Quantum Yang-Mills Theories - 2. Minkowski Axioms}

\author[1]{Min Chul Lee}
\email{min.lee@chch.ox.ac.uk}

\begin{abstract}
    This paper presents the axioms for a quantum Yang-Mills theory in the Minkowski spacetime. There are two routes of analytic continuation for the Schwinger functions, namely the Wightman functions and time-ordered products of field operators. We check consistency of these two axiom schemes and reproduce some well-known existing results, including the indefinite metric and Batalin–Vilkovisky formalisms.
\end{abstract}

\maketitle

Keywords : Quantum Yang-Mills Theories, Axiomatic Quantum Field Theory, Local Gauge Symmetry

\tableofcontents

\section{Introduction}
\label{sec:intro}

Analytic continuation from Euclidean to Minkowski metric.

Two types of such continuation are considered:
\begin{itemize}
 
      \item Vacuum expectation values\footnote{Also known as Wightman functions} of field operators in the sense of~\cite{StrWig64, OstSch73, OstSch75} as well as GNS reconstruction from them

      \item Time-ordered products of field operators in the sense of~\cite{time79} 
\end{itemize}
In both cases, local gauge action and gauge-invariant co-located products are considered. It is expected that our Schwinger functions are in one-to-one correspondence with time-ordered products, but not with Wightman functions. This is because Wightman functions are obtained only from Schwinger functions at non-coinciding points. However, reconstructed field operators from the Wightman functions are expected to yield the time-ordered products again, establishing certain consistency.

We will also make connection with algebraic quantum field theory and BV formalism.

\section{Notations}

Many of the notations here are the same as for the Euclidean axiom paper, except that the spacetime symmetry is modeled on the Poincar\'e group instead of the Euclidean group.

\begin{definition}
    Throughout this paper, the expression 
\[
C:=D
\]
means that $C$ is defined as $D$. The terms  ``Yang-Mills theory" and  ``Non-Abelian gauge theory" are used interchangeably.
\end{definition}

\begin{definition}
  \label{coordinate basic}
$\mu, \nu \in \{0,1,2,3\}$ denote the spacetime indices. On $\mathbb{R}^4$, let $\{e_0, e_1, e_2, e_3\}$ be the standard basis. Denote by $\{e^0, e^1, e^2, e^3\}$ the corresponding dual basis in the sense that
\[
e^\mu (e_\nu) = \delta^{\mu}_{\nu}.
\]
$\eta$ denotes the Minkowski metric on $\mathbb{R}^4$, so that
\[
\eta = \eta_{\mu \nu} e^\mu \otimes e^\nu
\]
with 
\[
\bigl(\eta_{\mu \nu} \bigr) = \begin{pmatrix}
-1 & 0 & 0 & 0 \\
0 & 1 & 0 & 0 \\
0 & 0 & 1 & 0 \\
0 & 0 & 0 & 1 \\.
\end{pmatrix}
\]
Throughout this paper, this pseudo-Riemannian structure is assumed on $\mathbb{R}^4$, except for Definition~\ref{local gauge g} below.

 $\undertilde{x}$ denotes a point in $\mathbb{R}^4$, which may be interpreted either as a vector (from the origin) with respect to the standard basis or a co-vector (from the origin) with respect to the dual basis. Its components are denoted in the former case as $(x^0,x^1,x^2,x^3)$ and in the latter case as $(x_0,x_1,x_2,x_3)$. We also use the notations $\undertilde{y}$, $\undertilde{w}$ and $\undertilde{z}$ with their components similarly to $\undertilde{x}$.
   
\end{definition}

\begin{definition}
For any $N \in \mathbb{N}$, $\S(\mathbb{R}^{4N})$ denotes the space of complex-valued Schwartz functions on $\mathbb{R}^{4N}$. $\S(\mathbb{R}^{4N})$ is a nuclear Fr\'echet space, with an example of the semi-norms explicitly presented in~\cite[p.86]{OstSch73}. Denote by $\S'(\mathbb{R}^{4N})$ the space of tempered distributions equipped with the strong dual topology. 
\end{definition}

\begin{definition}
\label{lie group and lie algebra}
     The gauge group $G$ denotes a simple compact Lie group, typically $SU(N)$ for $N \geq 2$. We fix a bi-invariant Riemannian metric on $G$ throughout this paper.
     
     The Lie algebra associated with $G$ is denoted as $\mathfrak{g}$ and we write its Lie bracket as $[ \, , \, ]$. Following~\cite[Chapter 15]{Wei96}, we fix a set of generators for $\mathfrak{g}$ 
     \[
     \bigl \{ t_\alpha \mid \alpha = 1,2, \cdots, \text{dim } \mathfrak{g} \bigr \}
     \]
     which are normalized in the sense that $\Tr(t_\alpha t_\beta)= \delta_{\alpha \beta}$. 
     
     %% 최소한 adjoint representation에서 이렇게 trace normalize가 된다고 가정...
     %임의의 representation에서 다 이렇게 되도록 할 수 있는지도 언젠가는 체크해보자...
The Peter-Weyl Theorem states that such $G$ may be realized as a matrix Lie group and the Lie bracket $[ \, , \, ]$ as the commutator of matrices. Such representations are assumed throughout this paper via specification of the matter and gauge fields in a given theory.

We also denote the linear mapping from $\mathfrak{g} \otimes \mathfrak{g}$ into $\mathbb{C}$ defined by $t_\alpha \otimes t_\beta \to \delta_{\alpha \beta}$ as $\Tr$, which is a slight abuse of the notation but will be clear from the context. Furthermore, denote as $[ \, , \, ]_{\otimes}$ the linear mapping on $\mathfrak{g} \otimes \mathfrak{g}$ defined by
\[
[ t_{\alpha} , t_\beta ]_{\otimes} := t_\alpha \otimes t_\beta - t_\beta \otimes t_\alpha
\]
   \end{definition}

\begin{definition}
\label{evaluation def}
  Let $\mathcal{M}$ be the space of smooth $\mathfrak{g}$-valued mappings on $\mathbb{R}^4$. For any $k \in \mathbb{N}$, denote by $[\mathcal{M}]^{\otimes k}$ the $k$-fold tensor product space with the topology as in~\cite[Chapter 43]{Treves67}. Moreover, let $\mathcal{M}_k$ be the space of smooth $\mathfrak{g}$-valued mappings on $\mathbb{R}^{4k}$.\footnote{Just like the Euclidean axiom paper, we may choose the alternative definition of $\mathcal{M}$ and $\mathcal{M}_k$ as the space of bounded $\mathfrak{g}$-valued smooth mappings with rapid decay of all Fr\'echet deriatives at infinity with respect to any operator norm, which will be more commonly used than the original one in the following sections.}

   Now, denote by $\mathscr{C}$ the unique continuous linear mapping from $[\mathcal{M}]^{\otimes k}$ to $\mathcal{M}_k$ given by the formula
   \[
   \mathscr{C}\Bigl( F_1 \otimes \cdots \otimes F_k \Bigr) :=  F_1(\undertilde{x_1}) \cdots  F_k(\undertilde{x_k}) 
   \]
   for $F_1 \cdots, F_k \in \mathcal{M}$.
\end{definition}

\begin{definition}
\label{colocated trace def}
  We denote by $\Tr_{nc}$ the composition mapping $\Tr \, \circ  \, \mathscr{C}$. That is,
   \begin{equation}
    \label{noncolocated trace}
       \Tr_{nc} :  F_1 \otimes \cdots \otimes F_k \in [\mathcal{M}]^{\otimes k} \to \Bigl( (\undertilde{x_1},\cdots,\undertilde{x_k}) \to \Tr\bigl(F_1(\undertilde{x_1}) \cdots  F_k(\undertilde{x_k})\bigr) \Bigr) \in C^\infty(\mathbb{R}^{4k}).
    \end{equation}

Moreover, we introduce the co-located trace
    \begin{equation}
    \label{colocated trace}
      \Tr_c :  F_1 \otimes \cdots \otimes F_k \in [\mathcal{M}]^{\otimes k} \to \Bigl( \undertilde{x} \to \Tr\bigl(F_1(\undertilde{x}) \cdots  F_k(\undertilde{x})\bigr) \Bigr) \in C^\infty(\mathbb{R}^4)
    \end{equation}
which again is uniquely extended to a continuous linear mapping. 
%%% MO에 올린 질문 체크!!!%%%
\end{definition}

 \begin{definition}
 \label{local gauge g}
    Let $g  : \mathbb{R}^4 \to G $ be a smooth mapping. Its differential $Dg$ is than a mapping of $\undertilde{x} \in \mathbb{R}^4$ such that
    \[
    D_{\undertilde{x}} g \in L\bigl(\mathbb{R}^4, T_{g(\undertilde{x})} G \bigr)
    \]
     Using the Euclidean norm $\lVert \cdot \rVert$ on $\mathbb{R}^4$ and the assumed bi-invariant Riemannian metric on $T_{g(\undertilde{x})} G$, we may define the operator norm $\lVert \cdot \rVert_{op}$ on $L\bigl(\mathbb{R}^4, T_{g(\undertilde{x})} G \bigr)$. Note that $\lVert \cdot \rVert_{op}$ does not depend on $x$ by construction. We say that $x \to D_{\undertilde{x}}g$ is rapidly decaying at infinity if
\[
\sup_{\undertilde{x} \in \mathbb{R}^4} \bigl(1 + \lVert \undertilde{x} \rVert \bigr)^n \lVert D_x g \rVert_{op} < \infty
\]
for all $n \in \mathbb{N}$. It is straightforward to generalize this notion of rapid decay at infinity to higher-order differentials of $g$.  

In order to address the local action of $G$ in the context of tempered distribution, we introduce the following group:
\[
\mathcal{G}:=\Bigl\{ g : \mathbb{R}^4 \to G \, \bigl \lvert \, g\text{ is smooth with the differential of each order rapidly decaying at infinity} \Bigr\}
\]   
where the group operation is defined point-wise. That is, $(g_1 g_2)(\undertilde{x}):=g_1(\undertilde{x}) g_2(\undertilde{x})$ and $g^{-1}(\undertilde{x}):=[g(\undertilde{x})]^{-1}$. 

Since $G$ is a compact Lie group, $\mathcal{G}$ is indeed a group under such operations. Moreover, it is not difficult to observe that components of any $g \in \mathcal{G}$ with respect to any representation of $G$ are smooth bounded functions on $\mathbb{R}^4$ whose partial derivatives are all rapidly decaying at infinity.
\end{definition}

\begin{definition}
\label{gauge subtlety}
    Let $A : \mathbb{R}^{4 \times 2} \to \mathbb{C}$ be a bounded smooth function whose partial derivatives of all orders are rapidly decreasing. We denote by $A \bigl \lvert_{\text{diag}}$ the function $\undertilde{x} \to A(\undertilde{x},\undertilde{x})$.\footnote{This definition has been motivated by $\mathcal{G}$ above and will be crucial for gauge invariance of co-located products of Yang-Mills curvature tensors.} 
\end{definition}

\begin{definition}
\label{gauge index def}
    We assume that the matter fields are partitioned into multiplets to furnish representations of the gauge group $G$ in a given theory. More specifically, let $\mathcal{R}$ be the labeling of all multiplets in the theory. Then, there exists a representation of $G$ for each $r \in \mathcal{R}$ such that the $r$-multiplet of matter fields are components with respect to a given (ordered) basis in the representation space. We denote the representation space by $V_r$ and the basis by $\{e_{k_r}\}$; here the index $k_r$ takes the value from $\{1, 2, \cdots, \text{dim } V_r\}$.
    
    We assume in addition that the adjoint\footnote{Here the term \textit{adjoint} refers to the dual spinor representation. For example, it denotes the Dirac adjoint in the case of a Dirac spinor.} of the matter fields in the $r$-multiplet forms a multiplet in the representation of $G$ dual to that of $r$ with respect to the dual basis of $\{e_{k_r}\}$. Such multiplet is labeled by $\overline{r}$. For any $g \in \mathcal{G}$, the notation $g_{(r)}$ is used to make it explicit the representation under which the values of $g$ are expressed.  
    
    Note that, by definition, $V_{\overline{r}}$ is the dual space of $V_r$ and $\{e_{k_{\overline{r}}}\}$ is the dual basis for $\{e_{k_r}\}$. We assume further that $\{e_{k_{\overline{r}}}\}$ is ordered in the same way as $\{e_{k_r}\}$ so that $\langle e_{k_{\overline{r}}}, e_{k_r} \rangle_{V_{\overline{r}} \times V_r}=\delta_{k_r k_{\overline{r}}}$ and the adjoint of the matter field component corresponding to $e_{k_r}$ is the component in the $\overline{r}$-multiplet corresponding to $e_{k_{\overline{r}}}$. We further assume the following completeness relation:
    \[
   \sum_{k_r, k'_{\overline{r}}} \delta_{k_r k'_{\overline{r}}}\bigl\langle \phi, e_{k_r} \bigr \rangle_{V_{\overline{r}} \times V_r } \bigl \langle e_{k'_{\overline{r}}} , v \bigr \rangle_{V_{\overline{r}} \times V_r } = \bigl \langle \phi , v \bigr \rangle_{V_{\overline{r}} \times V_r } \text{ for all } \phi \in V_{\overline{r}} \text{ and } v \in V_r
    \]

    For notational convenience, we may also use the symbols $V^*_r$ and $\{e^*_{k_r}\}$ to denote the dual space $V_{\overline{r}}$ and the dual basis $\{e_{k_{\overline{r}}}\}$ respectively. Note that $e^*_{k_r}=\sum_{k'_{\overline{r}}}\delta_{k_r k'_{\overline{r}}}e_{k'_{\overline{r}}}$.
\end{definition}

\begin{definition}
\label{spinor index def}
    For the spinor indices of matter fields, we modify~\cite[Section 6]{OstSch73}. All fields in each $r$-multiplet are assumed to be of the same spinor character in the Minkowski spacetime. As such, the index $\nu_r$ describes the spinor character of the fields in the $r$-multiplet. By construction, the index $\nu_{\overline{r}}$ corresponds to the representation of the restricted Lorentz group $SO^{+}(1,3)$ dual to that of $\nu_r$.
\end{definition}

\begin{definition}
    $\Psi_{k_r \nu_r}$ denotes the matter field which is the component of the basis element $e_{k_r}$ in the $r$-multiplet, so that the whole multiplet may be expressed as $\Psi_{v_{r}}=\sum_{k_r}\Psi_{k_r \nu_r}e_{k_r}$. With the notations from Definition~\ref{gauge index def} and~\ref{spinor index def}, the adjoint of the field $\Psi_{k_r \nu_r}$ is $\Psi_{k_{\overline{r}} \nu_{\overline{r}} }$ and vice versa.\footnote{As a concrete example, let us consider QCD~\cite[Section 18.7]{Wei96}, where quark fields and their Dirac adjoints are the matter fields. In this case, $\mathfrak{R}=\{ u, \overline{u}, c,\overline{c}, t, \overline{t}, d,\overline{d}, s,\overline{s},b,\overline{b} \}$ corresponds to the flavors, where $u$ denotes the $u$-quark \textit{field} while $\overline{u}$ is its Dirac adjoint, and similarly for other flavors. This is slightly different from the conventional meaning of $u$ and $\overline{u}$ as a \textit{particle} and anti-\textit{particle} respectively, but makes no essential change. With $r$ denoting any element of $\{ u,c,t,d,s,b\}$, $k_r$ takes three values (= colors), furnishing the fundamental representation $\bm{3}$ of $SU(3)$. $k_{\overline{r}}$ takes three values as well, furnishing the dual representation $\overline{\bm{3}}$.} 
\end{definition}

\begin{definition}
\label{euclidean rep}
   For $U,V \in SL(2,\mathbb{C})$, we denote by $R(U,V)$ the representation of $SO^{+}(1,3)$ corresponding to a given spinor index and $R=R(U,V)$ for the fundamental representation of $SO(4)$. This is identical to the notations in~\cite[p.102]{OstSch73}.
\end{definition}

\section{The Minkowski Axioms for a Quantum Yang-Mills theory}

\subsection{Axioms for Wightman functions}

\subsection{Axioms for Time-Ordered Operator Products}

In this subsection, we present the axioms for time-ordered products of field operators for a quantum Yang-Mills theory. As a consistency check, it is shown that the vacuum expectation values of the field operators here coincide with the Wightman functions in the previous subsection.

%% counterterm에 해당하는 field operator product 같은 게 있으려나?? 파악해보자..

\[
1+1=2 \tag{E0} \label{E0}
\]

A cross-reference to equation~\eq{E0}.

\section{Analytic Continuation and Reconstruction Theorems}

As described in the Introduction

\section{Relation to the Indefinite Metric Formalism}

In~\cite[Ch.10]{BoLoOkTo90}, a general framework for the state space of a Non-Abelian gauge theory is presented, with rigorous justification of related physics derived from this formalism. In this section, we reproduce most of the results of~\cite[Ch.10]{BoLoOkTo90} from our reconstruction theorem in the previous section.

One major element that we cannot reproduce is, however, the assumption of an initial underlying Hilbert space in~\cite[Ch.10]{BoLoOkTo90}. There is no guarantee based on our reconstruction theorem that the initial state space $H$ is a Hilbert space with the sesquilinear form $\langle \, , \rangle$ expressed in terms of the inner product as in~\cite[p.418, Eq. (10.2)]{BoLoOkTo90}. 

Nevertheless, we do not regard this as a critical defect, since it is emphasized that the sesquilinear form $\langle \, , \rangle$ (or \textit{indefinite scalar product} as stated in the reference) is the important part.

%5여기 내용을 완전히 reproduce하기는 힘들수도...특히 저 ch.10 보면 underlying Hilbert space를 gauge-dependent field operator들에 대해서도 전제로 하고 가는데...그건 너무 strong assumption 같음...

%%그러나 syemmtry breaking이나 goldstone boson, mass gap 등은 어떻게든 derive해야 할듯...

%%또한 algebra of observables 등에 대한 axiom들도 다 verify해야 할듯...(larger group vs little group 내용 좀 더 정확히 보자...)

\section{Relation to AQFT and BV formalism}

\section{Conclusion}
We are deeply grateful for Professors Arthur Jaffe,  Klaus Fredenhagen and Kasia Rejzner for their critical comments and insights.

\bibliographystyle{unsrt}
\bibliography{refs}

\end{document}